\documentclass[journal]{IEEEtran}
\usepackage{graphicx}
\usepackage{cite}
\usepackage{amsmath,amssymb,amsfonts}
\usepackage{algorithmic}
\usepackage{textcomp}
\usepackage{booktabs}
\usepackage{xspace}
\usepackage{svg}
\usepackage{amsmath}
\usepackage{footnote}
\usepackage{makecell}
\usepackage{multirow}
\usepackage{enumitem}
\usepackage{array}
\usepackage[nolist,nohyperlinks]{acronym}
\usepackage{hyperref}
\usepackage{subcaption}
\makesavenoteenv{tabular}
\usepackage{caption, subcaption}

\usepackage[acronym]{glossaries}
\acrodef{CRC}{colorectal cancer}
\acrodef{GI}{Gastrointestinal}
\acrodef{mIoU}{mean Intersection over Union}
\acrodef{ReLU}{Rectified Linear Unit}
\acrodef{MIS}{Minimally invasive surgery}
\acrodef{RAS}{Robot assisted surgery}
\acrodef{CADx}{Computer-Aided Detection}
\acrodef{CNN}{Convolutional Neural Network}
\acrodef{DNN}{Deep Neural Network}
\acrodef{DL}{Deep Learning}
\acrodef{ML}{Machine Learning}
\acrodef{RNN}{Residual Neural Network}
\acrodef{GAN}{Genearative Adversarial Network}
\acrodef{FPS}{Frame Per Second}
\acrodef{WCE}{Wireless Capsule Endoscopy}
\acrodef{CGAN}{Conditional Generative Adversarial Network}
\acrodef{MICCAI}{Medical Imaging Computing and Computer Assisted Intervention Society}
\acrodef{AI}{Artificial Intelligence}
\acrodef{VCE}{Video Capsule Endoscopy}
\acrodef{DSC}{Dice Coefficient}
\acrodef{SOTA}{state-of-the-art}
\acrodef{FPS}{Frame-per-second}
\begin{document}
\title{NanoNet: Real-Time Polyp Segmentation in Video Capsule Endoscopy and Colonoscopy}


\author{
\IEEEauthorblockN{Debesh Jha\IEEEauthorrefmark{1}\IEEEauthorrefmark{2}, 
Nikhil Kumar Tomar\IEEEauthorrefmark{1},
Sharib Ali\IEEEauthorrefmark{4}, 
Michael A. Riegler\IEEEauthorrefmark{1}, \\
H{\aa}vard D. Johansen\IEEEauthorrefmark{2},
Dag Johansen\IEEEauthorrefmark{2},
Thomas de Lange\IEEEauthorrefmark{5}\IEEEauthorrefmark{6}\IEEEauthorrefmark{7}\IEEEauthorrefmark{8},  
P{\aa}l Halvorsen\IEEEauthorrefmark{1}\IEEEauthorrefmark{3}
}
\vspace{3mm}

\IEEEauthorblockA{\IEEEauthorrefmark{1}SimulaMet, Norway \ \ 
\IEEEauthorrefmark{2}UiT The Arctic University of Norway, Norway \ \ 
\IEEEauthorrefmark{3}Oslo Metropolitan University, Norway \ \ 
\IEEEauthorrefmark{4}Institute of Biomedical Engineering, University of Oxford, Oxford, UK\\ 
\IEEEauthorrefmark{5}Department of Medical Research, Bærum Hospital, Norway \ \ 
\IEEEauthorrefmark{6}Augere Medical AS, Norway\\ 
\IEEEauthorrefmark{7}Medical Department, Sahlgrenska University Hospital-Mölndal Hospital, Sweden\\ 
\IEEEauthorrefmark{8}Department of Molecular and Clinical Medicine, Sahlgrenska Academy, University of Gothenburg, Sweden\\
}
}

\maketitle
\begin{abstract}
Deep learning in gastrointestinal endoscopy can assist to improve clinical performance and be helpful to assess lesions more accurately. To this extent, semantic segmentation methods that can perform automated real-time delineation of a region-of-interest, e.g., boundary identification of cancer or precancerous lesions, can benefit both diagnosis and interventions. However, accurate and real-time segmentation of endoscopic images is extremely challenging due to its high operator dependence and high-definition image quality. To utilize automated methods in clinical settings, it is crucial to design lightweight models with low latency such that they can be integrated with low-end endoscope hardware devices. In this work, we propose \textit{NanoNet}, a novel architecture for the segmentation of video capsule endoscopy and colonoscopy images. Our proposed architecture allows real-time performance and has higher segmentation accuracy compared to other more complex ones. We use video capsule endoscopy and standard colonoscopy datasets with polyps, and a dataset consisting of endoscopy biopsies and surgical instruments, to evaluate the effectiveness of our approach. Our experiments demonstrate the increased performance of our architecture in terms of a trade-off between model complexity, speed, model parameters, and metric performances. Moreover, the resulting models´ size is relatively tiny, with only nearly 36,000 parameters compared to traditional deep learning approaches having millions of parameters. 
\end{abstract}

\IEEEpeerreviewmaketitle

\section{Introduction}
\label{sec:introduction} 
 
\ac{GI} endoscopy is a widely used technique to diagnose and treat anomalies in the upper (esophagus, stomach, and duodenum) and the lower (large bowel and anus) GI tract. Among the other \ac{GI} tract organs, \ac{CRC} has the highest cancer incidences and mortality rate~\cite{sung2021global}. There are several \ac{CRC} screening options. Theses are usually divided into two categories, namely, invasive (visual examination-based test) and non-invasive based tests (stool, blood, and radiological test). \emph{Colonoscopy}, the gold standard for examining the large bowel (colon and rectum), is an invasive examination used to detect, observe, and remove abnormalities (such as polyps). It detects colorectal cancer with both high sensitivity and specificity. \emph{Sigmoidscopy} is another invasive test. \emph{Computed Tomography(CT) Colonoscopy}, \emph{Fecal Occult Blood Test (FOBT) Fecal Immunochemical Test (FIT)}, and \ac{VCE} are non-invasive tests. \ac{VCE} is a technology for capturing the video inside the \ac{GI} tract.  It has evolved as an important tool for detecting small bowel diseases~\cite{kornbluth2004video}.

\ac{DL} methods have made a significant breakthrough in several medical domain such as lung cancer detection~\cite{ardila2019end}, diabetic retinopathy progression~\cite{arcadu2019deep}, and obstructive hypertrophic cardiomyopathy detection~\cite{green2019machine}. It has provided new opportunities to solve challenges such as bleeding, light over/underexposure, smoke, and reflections~\cite{bodenstedt2018comparative}. However, \ac{DL} normally needs a large annotated dataset for the implementation of methods. It is difficult to obtain a labeled medical dataset. First, it needs collaborations with the hospitals.  For data collection, the doctors require approval from various authorities and patient consent. They need to set protocols for the collection, and the collected data must be anonymized and cleaned with the help of data engineers.  Domain experts must label raw data, and after labeling, the annotations must be done depending upon the need of the task. The whole process requires an significant amount of expert time and is costly. Additionally, it is an operator-dependent process. The quality of the data labeling and annotation depends on the expertise of the clinicians. Therefore, it is challenging to curate a larger dataset.

One way of solving the dataset issue is to create synthetic images using \ac{GAN}~\cite{goodfellow2014generative}. However, generated synthetic images may not always capture all the properties and characteristics of real endoscopic images. Consequently, the model may only learn to predict the properties from the synthetic images and may not perform well on a real endoscopic dataset. Another solution could be domain adaptation from a similar endoscopic dataset. However, we lack large publicly available labeled endoscopic datasets. Thus, a viable and compelling approach to solve the semantic segmentation task is to reuse ImageNet pre-trained encoders in the segmentation model~\cite{chen2017deeplab}. The predicted masks from the algorithm can provide reliable information to the endoscopic model.

A lightweight \ac{CNN} model can be essential for the development of real-time and efficient semantic segmentation methods. Usually, lightweight models are computationally efficient and require less memory. A smaller number of parameters makes the network less redundant. Lightweight \ac{CNN} models are mainly being deployed in mobile applications~\cite{kim2015compression}. A lightweight model can play a crucial role from a system perspective with a limited resource constraint for real-time prediction in clinics.  Consequently, we propose a novel architecture, NanoNet, optimized for faster inference and high accuracy. An extremely lightweight model with very few trainable parameters, faster inference, and higher performance would require less memory footprint to be incorporated with any devices. Therefore, we put forward this approach to address the challenges in endoscopy.  

The main contributions of this work include the following:
\begin{enumerate}
 \item We proposed a novel architecture, named NanoNet, to segment video capsule endoscopy and colonoscopy images in real-time with high accuracy. The proposed architecture is very lightweight, and the model size is smaller, requiring less computational cost. 

\item \ac{VCE} datasets are difficult to obtain with pixel-wise annotations.  In this context, we have annotated 55 polyps from the ``polyp" class of the Kvasir-Capsule dataset with the help of an expert gastroenterologist. We have made this dataset public and provided the benchmark. 

\item NanoNet achieves promising performance on the KvasirCapsule-SEG, Kvasir-SEG~\cite{jha2020kvasir}, 2020 Medico automatic polyp segmentation challenge~\cite{jha2020medico}, 2020 EndoTect challenge~\cite{hicksendotect}, and Kvasir-Instrument~\cite{jha2020kvasirInstrument} datasets. All experiments conform with \ac{SOTA} in terms of parameter uses (size), speed, computation, and performance metrics.   
    
 \item The model can be integrated with mobile and embedded devices because of fewer parameters used in the network. 

\end{enumerate}

\section{Related work}
\label{sec:relatedwork}

\subsection{Semantic segmentation of endoscopic images}
Semantic segmentation of endoscopic images has been a well-established topic in medical image segmentation. Earlier work mostly relied on the handcrafted descriptors for feature learning~\cite{karkanis2003computer,ameling2009texture}. The handcrafted features such as color, shape, texture, and edges were extracted and fed to the \ac{ML} classifier, which separates lesions from the background. However, the traditional \ac{ML} methods based on handcrafted features suffer from low performance~\cite{bernal2012towards}. The recent works on polyp segmentation using both video capsule endoscopy and colonoscopy mostly relied on \ac{DNN}~\cite{jia2019wireless,prasath2017polyp,tomar2021fanet,guo2020polyp,ali2021deep,fan2020pranet,jhacomprehensive}.

With the \ac{DNN} methods, there is progress in the performance for segmenting endoscopic images (for example, polyps). However, the network architectures are often complex and requires high-end GPUs for training, and is computationally expensive~\cite{jha2019resunet++,jha2020doubleUNet,fan2020pranet}. Additionally, real-time lesion segmentation has often been ignored. Although there is some recent initiation for the real-time detection of endoscopic images, they have mostly used private datasets~\cite{lee2020real,yamada2019development,poon2020ai} for the experimentation. It is difficult to compare the new methods on these datasets and extend the benchmark. Therefore, there is a need for a benchmark on publicly available datasets to minimize the research gap towards building a clinically relevant model. 

\subsection{Lightweight model}
There are few works in the literature that have proposed lightweight models for image segmentation. Ni et al.~\cite{ni2020barnet} presented a novel bilinear attention network-based approach with an adaptive receptive field for the segmentation of surgical instruments. Wang et al.~\cite{wang2019lednet} proposed a lightweight encoder-decoder network (LEDNet), an encoder-decoder network that uses ResNet50 in the encoder block and attention pyramidal network in the decoder block. Beheshti et al.~\cite{beheshti2020squeeze} proposed SqueezeNet. The architecture of the SqueezeNet is inspired by UNet~\cite{romera2017erfnet}. The proposed model obtained a 12$\times$ reduction in model size and showed efficient performance in multiplication accumulation (mac) and memory uses. \\

From the above-related work, we identify a need for a real-time polyp segmentation method. A real-time polyp segmentation method can be achieved by building a lightweight network architecture by designing an efficient network with blocks that require fewer parameters. A lower number of network parameters will reduce the network complexity, leading to real-time or faster inference. In this respect, we propose NanoNet, which uses a lightweight pre-trained network MobileNetV2~\cite{sandler2018mobilenetv2}, and simple convolutional blocks such as residual block and squeeze and excite block.

\begin{figure}[t!]
    \centering
    \includegraphics[clip, width = 1.0\columnwidth]{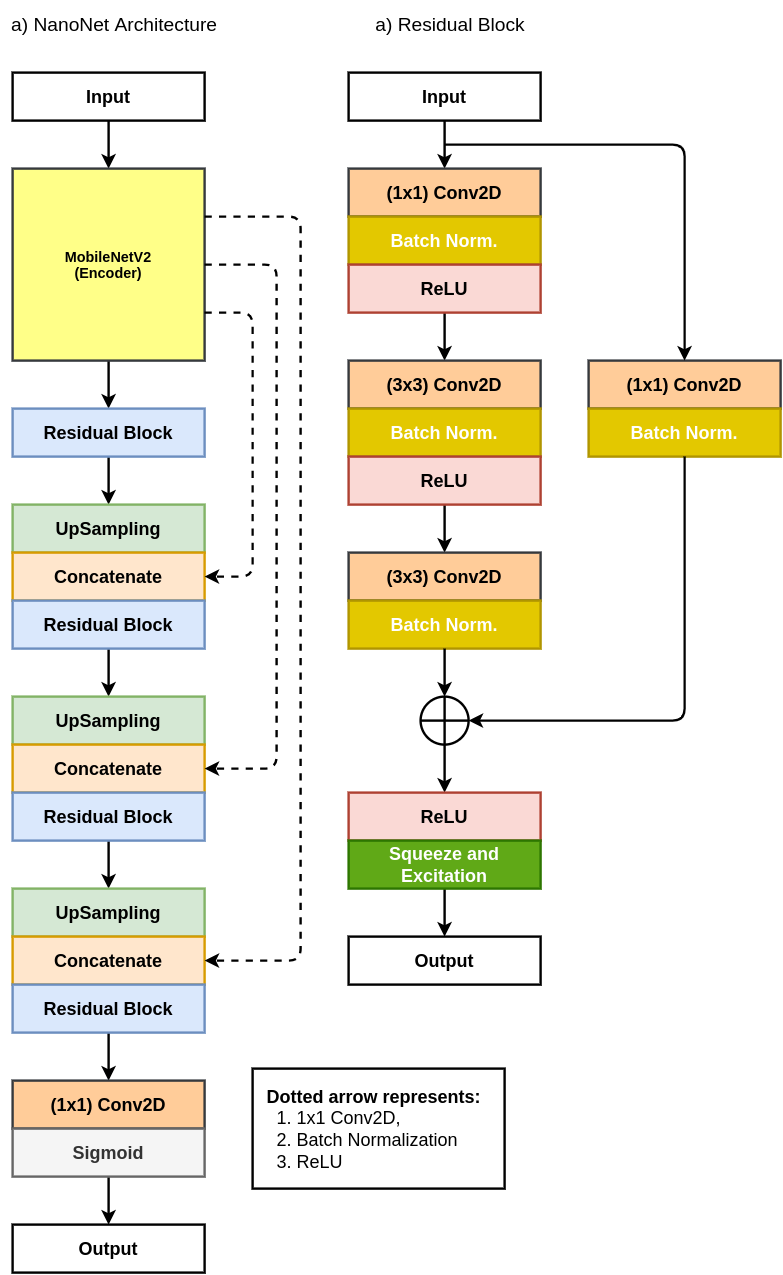}
    \caption{Overview of the proposed NanoNet architecture}
    \label{fig:nanonet}
\end{figure}
\section{Network architecture}
The architecture of NanoNet follows an encoder-decoder approach as shown in Figure~\ref{fig:nanonet}. As depicted in Figure~\ref{fig:nanonet}, the network architecture uses a pre-trained model as an encoder, followed by the three decoder blocks. Using pre-trained ImageNet~\cite{deng2009imagenet} models for transfer learning has become the best choice for many \ac{CNN} architectures~\cite{jha2020doubleUNet,chen2017deeplab}. It helps the model converge much faster and achieves high performance compared to the non-pre-trained model. The proposed architecture uses a MobileNetV2~\cite{sandler2018mobilenetv2} model pre-trained on the ImageNet~\cite{deng2009imagenet} dataset as the encoder. The decoder is built using a modified version of the residual block, which was initially introduced by He et al.~\cite{he2016deep}. The encoder is used to capture the required contextual information from the input, whereas the decoder is used to generate the final output by using the contextual information extracted by the encoder. 

\subsection{MobileNetV2}
The MobileNetV2~\cite{sandler2018mobilenetv2} is an architecture that is primarily designed for mobile and embedded devices. The architecture performed well on a variety of different datasets while maintaining high accuracy, despite having fewer parameters. The architecture of MobileNetV2 is based on the architecture of MobileNetV1, which uses depth-wise separable convolutions as the main building block. A depth-wise separable convolution consists of depth-wise convolution followed by a point-wise convolution. The MobileNetV2 introduces two main ideas: inverted residual block and linear bottleneck block~\cite{sandler2018mobilenetv2}. 

The inverted residual block is based on the bottleneck residual block as described in the~\cite{he2016deep}, which consists of three standard convolutions, which are $1\times1$, $3\times3$, and $1\times1$. Every convolution layer is followed by a \ac{ReLU} non-linearity. In the first $1\times1$ standard convolution, the number of feature channels are reduced, and in the last $1\times1$ standard convolution, the number of feature channels are expanded. After that, an element-wise addition with the identity mapping is performed. The inverted residual block also has three convolution layers: a $1\times1$ standard convolution, a $3\times3$ depth-wise convolution, and a $1\times1$ standard convolution. Every convolution has a \ac{ReLU} activation function. Here, the exact opposite of the bottleneck residual block is performed. The first $1\times1$ standard convolution expands the number of feature channels, and the last $1\times1$ standard convolution reduces the number of feature channels. Due to this opposite functionality, it is referred to as an inverted residual block. The linear bottleneck block is the same as the inverted residual block, except the last $1\times1$ standard convolution has a linear activation before an element-wise addition is performed with the identity mapping. 

\subsection{Modified Residual Block}
The original residual block uses two $3\times3$ standard convolutions, where the first convolution is followed by a batch-normalization and a \ac{ReLU} activation function. After that, the second convolution is followed only by a batch-normalization. An element-wise addition is performed between the output of the batch-normalization and the identity mapping, followed by another \ac{ReLU} activation function. An identity mapping consists of a $1\times1$ standard convolution and a batch-normalization over the original input. 

We have modified the residual block for our network. The modified residual block starts with a $1\times1$ convolution followed by a $3\times3$ convolution. In both of these convolutions, we reduce the number of filters by \(\frac{1}{4}\), which are then followed by the batch normalization and the \ac{ReLU} activation function. We have a $3\times3$ convolution with batch normalization.  Now, we perform an element-wise addition with the identity mapping. Finally, we apply a \ac{ReLU} activation function followed by the squeeze and excitation block. The squeeze and excitation block improves the quality of feature maps by increasing their sensitivity towards essential features. 

\begin{table*} [!t]
 \caption{Publicly available endoscopic datasets used in our experiments}
    \label{table:datasettable}
    \setlength{\tabcolsep}{6pt}
    \centering
    \begin{tabular}{p{4.5cm}|p{2cm}|p{3cm}|p{6cm}}
        \toprule
        Dataset & No. of Images & Imaging Type &Availability \\ 
        \midrule
        KvasirCapsule-SEG &55 &  Video capsule endoscopy & \url{https://www.dropbox.com/sh/hr46vieykbmvmkk/AAAs_V8ECG0wq51Fpw3rYU_5a?dl=0} \\ 
        Kvasir-SEG~\cite{jha2020kvasir}& 1000 & Colonoscopy &\url{https://datasets.simula.no/kvasir-seg/}\\ 
        2020 Medico automatic polyp segmentation challenge~\cite{jha2020medico} & 160$^\diamond$ &   Colonoscopy & \url{https://multimediaeval.github.io/editions/2020/tasks/medico/}\\  
        Endotect Challenge Dataset~\cite{hicksendotect} &200$^\diamond$  &  Colonoscopy &\url{https://endotect.com/}\\  
      Kvasir-Instrument~\cite{bernal2015wm}& 590 & Colonoscopy &\url{https://datasets.simula.no/kvasir-instrument/}\\  
        \bottomrule
        \multicolumn{3}{l}{$^\diamond${test images}}
\end{tabular}
\end{table*}	
\subsection{The NanoNet architecture}
Figure \ref{fig:nanonet} shows the block diagram of the NanoNet architecture. The NanoNet architecture starts with a pre-trained MobileNetV2 as an encoder followed by a decoder. There is a modified residual block between the encoder and the decoder, which acts like a bridge that connects the encoder and the decoder. In the first step, we feed the image data into the pre-trained encoder. The pre-trained encoder starts with a standard convolution with 32 feature channels, followed by the bottleneck layer with ReLU6 as the activation function. All the convolution operations use a standard $3\times3$ kernel size. The entire encoder network progressively downsamples the feature maps by using strided convolution and slowly increases the number of feature channels alternatively.  

The output from the pre-trained encoder passes through the modified residual block, which is fed to the decoder. Every step in the decoder uses a bilinear upsampling to increase the spatial dimension (height and width) of the input feature maps. After that, it is concatenated with the appropriate feature maps from the pre-trained encoder using the skip connections. These skip connections pass information that may be lost sometimes between the layers and are used to improve the quality of the feature maps. These concatenated feature maps are passed through the modified residual block, which further increases the generalization capacity of the decoder.  After the feature maps pass through all the three decoder block, the output of the last decoder block is fed to a  $1\times1$ convolution with a number of classes as the feature channels. This is followed by the sigmoid activation if it is a binary segmentation task, else we use the softmax activation function.

We have demonstrated three different NanoNet architectures: NanoNet-A, NanoNet-B, and NanoNet-C. Each architecture consists of different feature channels in its decoder block. NanoNet-A consists of  $32$, $64$ and $128$ feature channels. In NanoNet-B, the number of feature channels is reduced to $32$, $64$, and $96$. In NanoNet-C, these feature channels are further reduced to $16$, $24$, and $32$. The reduction in the number of feature channels leads to less trainable parameters, which simplifies the model complexity leading to a light-weight network.

\section{Experimental setup}
\label{sec:materials used}
In this section, we will describe the dataset, evaluation metrics, implementation details, and data augmentation techniques used. 

\begin{figure} [!t]
    \centering
    \includegraphics[height=4cm]{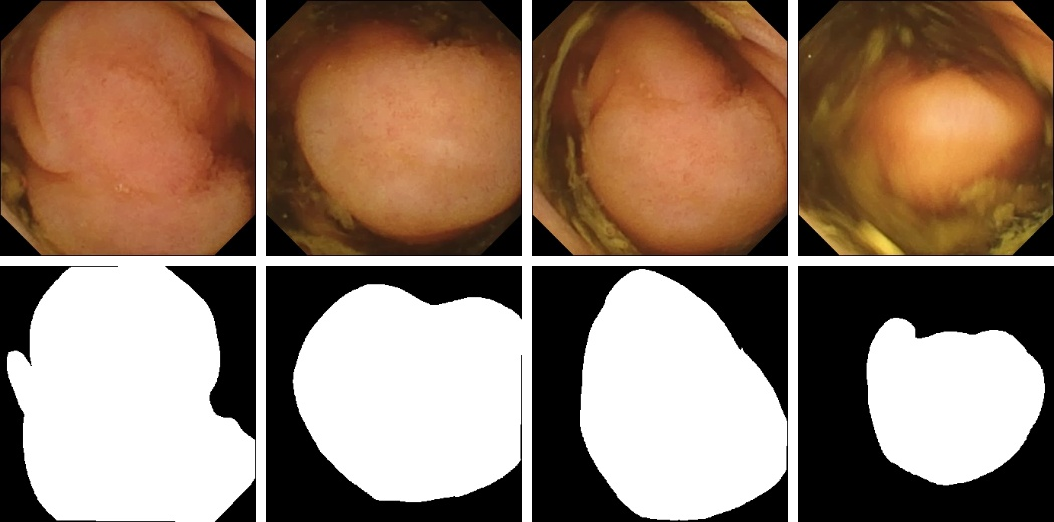}
    \caption{Polyps and corresponding masks from KvasirCapsule-SEG}  
    \label{fig:polypandmasks}
\end{figure}
\subsection{Datasets}
To address the polyp segmentation problem from video capsule endoscopy images, we have selected the polyp class from labelled images folder of the Kvasir-Capsule dataset~\cite{smedsrud2020kvasir} and annotated it with the help of an expert gastroenterologist. The  Kvasir-Capsule is an open-access dataset that contains 13 classes of labelled anomalies and findings. It only includes 55 polyp frames out of 44,228 medically verified video capsule frames present in the Kvasir-Capsule. We have annotated the polyp class of Kvasir-Capsule and generated corresponding ground truth masks. Examples of polyps and their corresponding masks from KvasirCapsule-SEG can be found in Figure~\ref{fig:polypandmasks}. Furthermore, we also provide bounding box information to be used for video capsule endoscopy detection and localization tasks. The Kvasir-Capsule can be downloaded from here~\footnote{\url{https://osf.io/dv2ag/}} and KvasirCapsule-SEG can be downloaded from here~\footnote{\url{https://www.dropbox.com/sh/hr46vieykbmvmkk/AAAs_V8ECG0wq51Fpw3rYU_5a?dl=0}}.

 Table~\ref{table:datasettable} shows the detailed information about the open imaging dataset used in our experiments. Each of the datasets presented in Table~\ref{table:datasettable} also has the corresponding ground truth. The link for each of the datasets is provided in the table. The standard setting for the ``Medico automatic polyp segmentation challenge" and ``Endotect challenge" is that they use the Kvasir-SEG for training. The challenge organizers have provided unseen 160 images in the ``Medico automatic polyp segmentation challenge" and released 200 images in the ``Endotect challenge" to test the participant's approaches. For the Kvasir-instrument dataset, we experimented with the official split provided by the organizers. The detail explanation of these datasets and the baseline results can be found in~\cite{jha2020kvasir,jha2020kvasirInstrument,jha2020medico,hicksendotect}.

\subsection{Evaluation metrics}
For the evaluation of our model, we have chosen standard computer vision metrics such as \ac{DSC}, \ac{mIoU}, Precision, Recall, Specificity, Accuracy, and \ac{FPS}. More explanation of these metrics can be found in~\cite{jha2020kvasir,jha2020kvasirInstrument,jha2020medico,hicksendotect}.

\begin{table*}[!t]
\centering
\caption{Performance evaluation of the proposed networks and recent \ac{SOTA} methods on KvasirCapsule-SEG}
\footnotesize
\begin{tabular}{l|l|l|l|l|l|l|l|l}
\toprule
\textbf{Method} &\textbf{Parameters} & \textbf{DSC} & \textbf{mIoU} & \textbf{Recall} & \textbf{Precision}& \textbf{F2} & \textbf{Accuracy} &\textbf{FPS} \\ \hline
ResUNet (GRSL'18)~\cite{zhang2018road}&8,227,393 &\textbf{0.9532} &\textbf{0.9137} &\textbf{0.9785} &\textbf{0.9325} &\textbf{0.9677} &\textbf{0.9386} &17.96 \\ 

ResUNet++ (ISM'19)\cite{jha2019resunet++} & 4,070,385 & 0.9499 & 0.9087 & 0.9762 & 0.9296 & 0.9648 & 0.9334 & 15.39\\

NanoNet-A (Ours) &235,425 &0.9493 &0.9059 &0.9693 & \textbf{0.9325} &0.9609 &0.9351 &28.35  \\ 
NanoNet-B (Ours)&132,049 &0.9474 &0.9028 &0.9682 &0.9308 &0.9593 &0.9324 &27.39 \\ 
NanoNet-C (Ours)&36,561  &0.9465 &0.9021 &0.9754 &0.9238 &0.9629 &0.9297 &\textbf{29.48}\\
\bottomrule
\end{tabular}
\label{tab:resultKvasirCapsule-SEG}
\end{table*}
\begin{table*}[!t]
\centering  
\caption{Performance evaluation of the proposed networks and recent \ac{SOTA} methods on Kvasir-SEG\cite{jha2020kvasir}}
\footnotesize
\begin{tabular}{l|l|l|l|l|l|l|l|l}
\toprule
\textbf{Method} &\textbf{Parameters} & \textbf{DSC}  & \textbf{mIoU} & \textbf{Recall} & \textbf{Precision} & \textbf{F2} & \textbf{Accuracy} & \textbf{FPS} \\ \hline
ResUNet (GRSL'18)~\cite{zhang2018road} & 8,227,393 & 0.7203 & 0.6106 & 0.7602 & 0.7624 & 0.7327 & 0.9251 & 17.72 \\ 

ResUNet++ (ISM'19)~\cite{jha2019resunet++} & 4,070,385 & 0.7310 & 0.6363 & 0.7925 & 0.7932 & 0.7478 & 0.9223 & 19.79 \\

NanoNet-A (Ours)& 235,425  & \textbf{0.8227} & \textbf{0.7282} & \textbf{0.8588} & \textbf{0.8367} & \textbf{0.8354}& \textbf{0.9456} & 26.13  \\ 
NanoNet-B (Ours)& 132,049  & 0.7860 & 0.6799 & 0.8392 & 0.8004 & 0.8067 & 0.9365 & 29.73\\ 
NanoNet-C (Ours)& 36,561   & 0.7494 & 0.6360 & 0.8081 & 0.7738 & 0.7719 & 0.9290 & \textbf{32.17}\\ 
\bottomrule
\end{tabular}
\label{tab:resultKvasir-SEG}

\end{table*}
\begin{table*}[!t]
\centering
\caption{Performance evaluation of the proposed networks and recent \ac{SOTA} methods on the Medico 2020 dataset~\cite{jha2020medico}}
\footnotesize
\begin{tabular}{l|l|l|l|l|l|l|l|l}
\toprule
\textbf{Method}&\textbf{Parameters} & \textbf{DSC} & \textbf{mIoU} & \textbf{Recall} & \textbf{Precision}& \textbf{F2} & \textbf{Accuracy} &\textbf{FPS} \\ \hline


ResUNet (GRSL'18)~\cite{zhang2018road} & 8,227,393 &0.6846 &0.5599 &0.7235 &0.7236 &0.6961 &\textbf{0.9231} & 18.54 \\ 

ResUNet++ (ISM'19)~\cite{jha2019resunet++} & 4,070,385 &0.6925 &0.5849 &0.8249 &0.6840 &0.7434 &0.8995 & 19.47 \\ 

NanoNet-A (Ours)&235,425 &0.7364 &\textbf{0.6319}  &\textbf{0.8566} &0.7310 &\textbf{0.7804} &0.9166 &28.07\\ 
NanoNet-B (Ours)&132,049  &\textbf{0.7378} &0.6247 &0.8283  &\textbf{0.7373} &0.7685 &0.9223 &29.04\\ 
NanoNet-C (Ours)&36,651 & 0.7070 &0.5866  &0.8095  &0.7089 &0.7432 &0.9148 &\textbf{32.66}\\ 
\bottomrule
\end{tabular}
\label{tab:resultMedico2020}
\end{table*}
%
\begin{table*}[!t]
\centering
\caption{Performance evaluation of the proposed networks and recent \ac{SOTA} methods on the Endotect 2020 dataset~\cite{hicksendotect}}
\footnotesize
\begin{tabular}{l|l|l|l|l|l|l|l|l}
\toprule
\textbf{Method} &\textbf{Parameters} & \textbf{DSC} & \textbf{mIoU} & \textbf{Recall} & \textbf{Precision}& \textbf{F2} & \textbf{Accuracy} &\textbf{FPS} \\ \hline


ResUNet (GRSL'18)~\cite{ronneberger2015u} & 8,227,393 &0.6640 &0.5408 &0.7510 &0.6841 &0.6943 &0.9075 & 26.55 \\ 

ResUNet++ (ISM'19)~\cite{jha2019resunet++} & 4,070,385 &0.6940 &0.5838 &\textbf{0.8797} &0.6591 &0.7597 &0.8841  &18.58 \\

NanoNet-A (Ours) & 235,425 &\textbf{0.7508} &\textbf{0.6466} &0.8238 &\textbf{0.7744} &\textbf{0.7773} &\textbf{0.9255} & 27.19 \\ 
NanoNet-B (Ours) & 132,049 &0.7362 &0.6238 &0.8109 &0.7532 &0.7646 &0.9252 & 29.91\\ 
NanoNet-C (Ours)&  36,651 &0.7001 &0.5792 &0.8000 &0.7159 &0.7380 &0.9091 & \textbf{32.98}\\ 
\bottomrule
\end{tabular}
\label{tab:resultEndotect2020}
\end{table*}
%
\begin{table*}[!t]
\centering
\caption{Performance evaluation of the proposed networks and recent \ac{SOTA} methods on Kvasir-Instrument~\cite{jha2020kvasirInstrument}}
\footnotesize
\begin{tabular}{l|l|l|l|l|l|l|l|l}
\toprule
\textbf{Method}&\textbf{Parameters} & \textbf{DSC} & \textbf{mIoU} & \textbf{Recall} & \textbf{Precision}& \textbf{F2} & \textbf{Accuracy} &\textbf{FPS} \\ \hline

UNet (Baseline)~\cite{ronneberger2015u}&- & 0.9158 & 0.8578 & \textbf{0.9487} & 0.8998 & \textbf{0.9320} & 0.9864 & 20.46\\ 
DoubleUNet (Baseline)~\cite{jha2020doubleUNet}& - & 0.9038 & 0.8430 & 0.9275 & 0.8966 & 0.9147 & 0.9838 & 10.00 \\ 

ResUNet++ (ISM'19)~\cite{jha2019resunet++}& 4,070,385 & 0.9140 & 0.8635 & 0.9103 & 0.9348 & 0.9140 & 0.9866 & 17.87 \\

NanoNet-A (Ours) & 235,425 & 0.9251 & 0.8768 & 0.9142 & \textbf{0.9540} & 0.9251 & \textbf{0.9887} & 28.00 \\ 
NanoNet-B (Ours) & 132,049 & \textbf{0.9284} & \textbf{0.8790} & 0.9205 & 0.9482 & 0.9284 & 0.9875 & 29.82 \\ 
NanoNet-C (Ours)& 36,561  & 0.9139 & 0.8600 & 0.9037 & 0.9452 & 0.9139 & 0.9863 & \textbf{32.18} \\ 
\bottomrule
\end{tabular}
\label{tab:resultKvasir-Instrument}
\end{table*}

\subsection{Implementation details}
We have implemented the NanoNet using Keras\footnote{https://keras.io/} with TensorFlow~\cite{abadi2016tensorflow} as backend. The experiments were run on the Experimental Infrastructure for Exploration of Exascale Computing (eX3), NVIDIA DGX-2 machine. The code implementation of NanoNet can be found here\footnote{\url{https://github.com/DebeshJha/NanoNet}}.  As the model has very few low trainable parameters, we have set a batch size of 16. We have resized the dataset images to $256 \times 256$ pixels for better utilization of the GPU, and it also helps to reduce the training time. The model is trained on 200 epochs with the Nadam optimizer~\cite{dozat2016incorporating} and dice coefficient as the loss function. The learning rate for the optimizer is set to 1e$^{-4}$. We prefer to choose a low learning rate to update the parameters slowly and carefully. The learning rate is reduced by a factor of 0.1 when the validation loss does not decrease in $10$ consecutive epochs.  It helps to improve model performance. Additionally, we have used an early stopping mechanism to prevent over-fitting.

\subsection{Data augmentation}
We use data-augmentation on the training set to increase diversity and to improve the generalization of our model. Data augmentation techniques such as random cropping, random rotation, horizontal flipping, vertical flipping, grid distortion, and many more are used. We have used an offline data augmentation technique. The validation and testing set is not augmented and is directly resized into $256\times 256$.

\begin{figure}[!t]
    \centering
    \includegraphics [clip, width=\columnwidth]{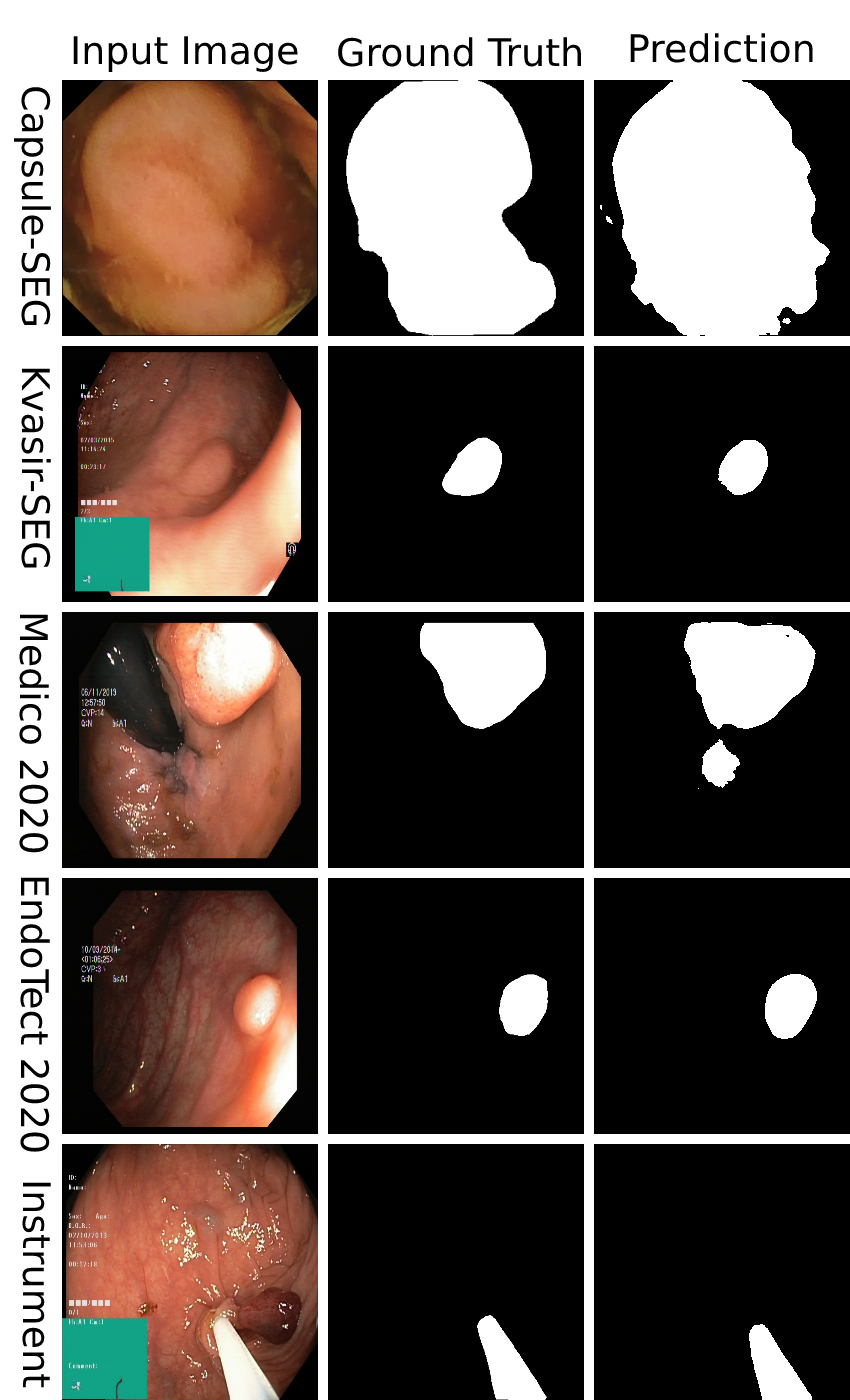}
    \caption{Qualitative results of NanoNet-A on five different datasets}
    \label{fig:qualitativeresultsa}
\end{figure}

\section{Result and Discussion}
\label{sec:results}
In this section, we provide the experimental results for the segmentation task of the endoscopic image dataset. For the evaluation, we have used performance metrics such as \ac{DSC} and \ac{mIoU}, and \ac{FPS} as the main evaluation metrics. We also calculate recall, precision, F2, and overall accuracy to support a complete set of metrics. Table~\ref{tab:resultKvasirCapsule-SEG}, Table~\ref{tab:resultKvasir-SEG}, Table~\ref{tab:resultMedico2020}, Table~\ref{tab:resultEndotect2020}, and Table~\ref{tab:resultKvasir-Instrument} show the results of the NanoNet model experiments using different parameters. The results are compared with the recent \ac{SOTA} computer vision methods.

The quantitative results in these tables show that NanoNet consistently outperforms or performs nearly equal to its competitors in terms of performance. The quantitative results also show that NanoNet can produce real-time segmentation (i.e., produces at least close to 30 \ac{FPS} for each dataset present in the Tables). This is one of the major contributions of the work. The other strength of the work lies in the parameter use. From Table~\ref{tab:resultKvasirCapsule-SEG}, we can observe that the best performing NanoNet (i.e., NanoNet-A) uses nearly 35 times less parameters as ResUNet~\cite{zhang2018road}. Similarly, NanoNet-C uses  225 times less parameters as compared to that of ResUNet and also produces better \ac{DSC}, \ac{mIoU} and \ac{FPS} with the Kvasir-SEG. 

The qualitative results are displayed in Figure \ref{fig:qualitativeresultsa}. The first, second, and third columns show the image, ground truth, and prediction masks, respectively. Similarly, the name of the dataset is provided on the left side. One example image for each dataset is shown.  The qualitative results with diversified classes of medical datasets show that NanoNet can produce accurate segmentation results with different types of lesions (polyps) and therapeutic tools. The example images and the prediction also show that NanoNet produces good segmentation masks for large, medium, and small polyps (see Figure \ref{fig:qualitativeresultsa}). From the qualitative results,  we can derive and conclude that NanoNet produces good results with small-sized polyps but produces over-segmentation for the large-sized lesions upon detail dissection. For future work, one could create a specific dataset consisting of a set of small and large-sized polyps to explore this further.

From both evaluation metrics and qualitative results, the improvement is remarkable.  Thus, the proposed NanoNet architecture is simple, compact, and provides a robust solution for real-time applications, as it produces satisfactory performance despite having fewer parameters.

\section{Conclusion}
\label{sec:conclusion}
In this paper, we proposed a novel lightweight architecture for real-time video capsule endoscopy and colonoscopy image segmentation. The proposed NanoNet architecture utilizes a pre-trained MobileNetV2 model and a modified residual block. The depthwise separable convolution is the main building block of the network and allows the model to achieve high performance with minuscule trainable parameters. The experimental results on varied endoscopy datasets demonstrate the strength of our model compared to state-of-the-art models with respect to their speed and performance. The presented model has the potential to enable easier roll out of deep learning models in clinical systems due to fewer parameters, competitive accuracy, and low-latency. In addition, the model does not require any sort of initialization, post-processing, or temporal regularization, considered as another strength of this work. In the future, we will design an encoder lighter than the currently used pre-trained MobilNetV2. Moreover, we aspire to utilize the currently built segmentation module in the clinic and study the efficacy of our designed model.

 \section*{Acknowledgment}
The research is partially funded by the PRIVATON project (263248) and the Autocap project (282315) from the Research Council of Norway (RCN). Our experiments were performed on the Experimental Infrastructure for Exploration of Exascale Computing (eX3) system, which is financially supported by RCN under contract 270053.

\bibliographystyle{IEEEtran}
\bibliography{references} 

\begin{thebibliography}{10}
\providecommand{\url}[1]{#1}
\csname url@samestyle\endcsname
\providecommand{\newblock}{\relax}
\providecommand{\bibinfo}[2]{#2}
\providecommand{\BIBentrySTDinterwordspacing}{\spaceskip=0pt\relax}
\providecommand{\BIBentryALTinterwordstretchfactor}{4}
\providecommand{\BIBentryALTinterwordspacing}{\spaceskip=\fontdimen2\font plus
\BIBentryALTinterwordstretchfactor\fontdimen3\font minus
  \fontdimen4\font\relax}
\providecommand{\BIBforeignlanguage}[2]{{%
\expandafter\ifx\csname l@#1\endcsname\relax
\typeout{** WARNING: IEEEtran.bst: No hyphenation pattern has been}%
\typeout{** loaded for the language `#1'. Using the pattern for}%
\typeout{** the default language instead.}%
\else
\language=\csname l@#1\endcsname
\fi
#2}}
\providecommand{\BIBdecl}{\relax}
\BIBdecl

\bibitem{sung2021global}
H.~Sung \emph{et~al.}, ``Global cancer statistics 2020: Globocan estimates of
  incidence and mortality worldwide for 36 cancers in 185 countries,''
  \emph{CA: a cancer journal for clinicians}, 2021.

\bibitem{kornbluth2004video}
A.~Kornbluth, P.~Legnani, and B.~S. Lewis, ``Video capsule endoscopy in
  inflammatory bowel disease: past, present, and future,'' \emph{Inflammatory
  Bowel Diseases}, vol.~10, no.~3, pp. 278--285, 2004.

\bibitem{ardila2019end}
D.~Ardila \emph{et~al.}, ``End-to-end lung cancer screening with
  three-dimensional deep learning on low-dose chest computed tomography,''
  \emph{Nature medicine}, vol.~25, no.~6, pp. 954--961, 2019.

\bibitem{arcadu2019deep}
F.~Arcadu \emph{et~al.}, ``Deep learning algorithm predicts diabetic
  retinopathy progression in individual patients,'' \emph{NPJ digital
  medicine}, vol.~2, no.~1, pp. 1--9, 2019.

\bibitem{green2019machine}
E.~M. Green \emph{et~al.}, ``Machine learning detection of obstructive
  hypertrophic cardiomyopathy using a wearable biosensor,'' \emph{NPJ digital
  medicine}, vol.~2, no.~1, pp. 1--4, 2019.

\bibitem{bodenstedt2018comparative}
S.~Bodenstedt \emph{et~al.}, ``Comparative evaluation of instrument
  segmentation and tracking methods in minimally invasive surgery,''
  \emph{arXiv preprint arXiv:1805.02475}, 2018.

\bibitem{goodfellow2014generative}
I.~J. Goodfellow \emph{et~al.}, ``Generative adversarial networks,''
  \emph{arXiv preprint arXiv:1406.2661}, 2014.

\bibitem{chen2017deeplab}
L.-C. Chen, G.~Papandreou, I.~Kokkinos, K.~Murphy, and A.~L. Yuille, ``Deeplab:
  Semantic image segmentation with deep convolutional nets, atrous convolution,
  and fully connected crfs,'' \emph{IEEE transactions on pattern analysis and
  machine intelligence}, vol.~40, no.~4, pp. 834--848, 2017.

\bibitem{kim2015compression}
Y.-D. Kim \emph{et~al.}, ``Compression of deep convolutional neural networks
  for fast and low power mobile applications,'' \emph{arXiv preprint
  arXiv:1511.06530}, 2015.

\bibitem{jha2020kvasir}
D.~Jha \emph{et~al.}, ``Kvasir-seg: A segmented polyp dataset,'' in \emph{Proc.
  of International Conference on Multimedia Modeling (MMM)}, 2020, pp.
  451--462.

\bibitem{jha2020medico}
D.~Jha, S.~A. Hicks, K.~Emanuelsen, H.~Johansen, D.~Johansen, T.~de~Lange,
  M.~A. Riegler, and P.~Halvorsen, ``Medico multimedia task at mediaeval 2020:
  Automatic polyp segmentation,'' in \emph{CEUR Proceedings of MediaEval
  Workshop}, 2020.

\bibitem{hicksendotect}
S.~A. Hicks \emph{et~al.}, ``The endotect 2020 challenge: Evaluation and
  comparison of classification, segmentation and inference time for
  endoscopy,'' in \emph{Proceedings of ICPR 2020 Workshops and Challenges},
  2020.

\bibitem{jha2020kvasirInstrument}
D.~Jha \emph{et~al.}, ``Kvasir-instrument: Diagnostic and therapeutic tool
  segmentation dataset in gastrointestinal endoscopy,'' in \emph{Proc. of
  Multimedia Modeling (MMM)}, 2021.

\bibitem{karkanis2003computer}
S.~A. Karkanis, D.~K. Iakovidis, D.~E. Maroulis, D.~A. Karras, and M.~Tzivras,
  ``Computer-aided tumor detection in endoscopic video using color wavelet
  features,'' \emph{IEEE transactions on information technology in
  biomedicine}, vol.~7, no.~3, pp. 141--152, 2003.

\bibitem{ameling2009texture}
S.~Ameling, S.~Wirth, D.~Paulus, G.~Lacey, and F.~Vilarino, ``Texture-based
  polyp detection in colonoscopy,'' in \emph{Bildverarbeitung f{\"u}r die
  Medizin 2009}, 2009, pp. 346--350.

\bibitem{bernal2012towards}
J.~Bernal, J.~S{\'a}nchez, and F.~Vilarino, ``Towards automatic polyp detection
  with a polyp appearance model,'' \emph{Pattern Recognition}, vol.~45, no.~9,
  pp. 3166--3182, 2012.

\bibitem{jia2019wireless}
X.~Jia, X.~Xing, Y.~Yuan, L.~Xing, and M.~Q.-H. Meng, ``Wireless capsule
  endoscopy: A new tool for cancer screening in the colon with
  deep-learning-based polyp recognition,'' \emph{Proceedings of the IEEE}, vol.
  108, no.~1, pp. 178--197, 2019.

\bibitem{prasath2017polyp}
V.~Prasath, ``Polyp detection and segmentation from video capsule endoscopy: A
  review,'' \emph{Journal of Imaging}, vol.~3, no.~1, p.~1, 2017.

\bibitem{tomar2021fanet}
N.~K. Tomar \emph{et~al.}, ``Fanet: A feedback attention network for improved
  biomedical image segmentation,'' \emph{arXiv preprint arXiv:2103.17235},
  2021.

\bibitem{guo2020polyp}
Y.~Guo, J.~Bernal, and B.~J~Matuszewski, ``Polyp segmentation with fully
  convolutional deep neural networks—extended evaluation study,''
  \emph{Journal of Imaging}, vol.~6, no.~7, p.~69, 2020.

\bibitem{ali2021deep}
S.~Ali \emph{et~al.}, ``Deep learning for detection and segmentation of
  artefact and disease instances in gastrointestinal endoscopy,'' \emph{Medical
  Image Analysis}, p. 102002, 2021.

\bibitem{fan2020pranet}
D.-P. Fan \emph{et~al.}, ``Pranet: Parallel reverse attention network for polyp
  segmentation,'' in \emph{Proc. of International Conference on Medical Image
  Computing and Computer-Assisted Intervention (MICCAI)}, 2020, pp. 263--273.

\bibitem{jhacomprehensive}
D.~Jha \emph{et~al.}, ``{A Comprehensive Study on Colorectal Polyp Segmentation
  with {ResUNet++}, Conditional Random Field and Test-Time Augmentation},''
  \emph{IEEE Journal of Biomedical and Health Informatics}.

\bibitem{jha2019resunet++}
D.~Jha, P.~H. Smedsrud, M.~A. Riegler, D.~Johansen, T.~De~Lange, P.~Halvorsen,
  and H.~D. Johansen, ``{ResUNet++:} {An Advanced Architecture for Medical
  Image Segmentation},'' in \emph{Proc. of IEEE International Symposium on
  Multimedia (ISM)}, 2019, pp. 225--2255.

\bibitem{jha2020doubleUNet}
D.~Jha, , M.~A. Riegler, D.~Johansen, P.~Halvorsen, and H.~D. Johansen,
  ``{DoubleU-Net:} {A Deep Convolutional Neural Network for Medical Image
  Segmentation},'' in \emph{Proc. of International Conference on Multimedia
  Modeling (MMM)}, 2020, pp. 451--462.

\bibitem{lee2020real}
J.~Y.~o. Lee, ``Real-time detection of colon polyps during colonoscopy using
  deep learning: systematic validation with four independent datasets,''
  \emph{Scientific reports}, vol.~10, no.~1, pp. 1--9, 2020.

\bibitem{yamada2019development}
M.~Yamada \emph{et~al.}, ``Development of a real-time endoscopic image
  diagnosis support system using deep learning technology in colonoscopy,''
  \emph{Scientific reports}, vol.~9, no.~1, pp. 1--9, 2019.

\bibitem{poon2020ai}
C.~C. Poon \emph{et~al.}, ``Ai-doscopist: a real-time deep-learning-based
  algorithm for localising polyps in colonoscopy videos with edge computing
  devices,'' \emph{NPJ Digital Medicine}, vol.~3, no.~1, pp. 1--8, 2020.

\bibitem{ni2020barnet}
Z.-L. Ni \emph{et~al.}, ``Barnet: Bilinear attention network with adaptive
  receptive field for surgical instrument segmentation,'' \emph{arXiv preprint
  arXiv:2001.07093}, 2020.

\bibitem{wang2019lednet}
Y.~Wang \emph{et~al.}, ``Lednet: A lightweight encoder-decoder network for
  real-time semantic segmentation,'' in \emph{Proc. of IEEE International
  Conference on Image Processing (ICIP)}, 2019, pp. 1860--1864.

\bibitem{beheshti2020squeeze}
N.~Beheshti and L.~Johnsson, ``Squeeze u-net: A memory and energy efficient
  image segmentation network,'' in \emph{Proc. of IEEE/CVF Conference on
  Computer Vision and Pattern Recognition (CVPR) Workshops}, 2020, pp.
  364--365.

\bibitem{romera2017erfnet}
E.~Romera, J.~M. Alvarez, L.~M. Bergasa, and R.~Arroyo, ``Erfnet: Efficient
  residual factorized convnet for real-time semantic segmentation,'' \emph{IEEE
  Transactions on Intelligent Transportation Systems}, vol.~19, no.~1, pp.
  263--272, 2017.

\bibitem{sandler2018mobilenetv2}
M.~Sandler, A.~Howard, M.~Zhu, A.~Zhmoginov, and L.-C. Chen, ``Mobilenetv2:
  Inverted residuals and linear bottlenecks,'' in \emph{Proc. of IEEE
  conference on computer vision and pattern recognition}, 2018, pp. 4510--4520.

\bibitem{deng2009imagenet}
J.~Deng \emph{et~al.}, ``Imagenet: A large-scale hierarchical image database,''
  in \emph{Proc. of IEEE conference on computer vision and pattern recognition
  (CVPR)}, 2009, pp. 248--255.

\bibitem{he2016deep}
K.~He, X.~Zhang, S.~Ren, and J.~Sun, ``Deep residual learning for image
  recognition,'' in \emph{Proc. of the IEEE conference on computer vision and
  pattern recognition (CVPR)}, 2016, pp. 770--778.

\bibitem{bernal2015wm}
J.~Bernal \emph{et~al.}, ``Wm-dova maps for accurate polyp highlighting in
  colonoscopy: Validation vs. saliency maps from physicians,''
  \emph{Computerized Medical Imaging and Graphics}, vol.~43, pp. 99--111, 2015.

\bibitem{smedsrud2020kvasir}
P.~H. Smedsrud \emph{et~al.}, ``Kvasir-capsule, a video capsule endoscopy
  dataset,'' \emph{Springer Nature Scientific Data}, 2021.

\bibitem{zhang2018road}
Z.~Zhang, Q.~Liu, and Y.~Wang, ``Road extraction by deep residual u-net,''
  \emph{IEEE Geoscience and Remote Sensing Letters}, vol.~15, no.~5, pp.
  749--753, 2018.

\bibitem{ronneberger2015u}
O.~Ronneberger, P.~Fischer, and T.~Brox, ``U-net: Convolutional networks for
  biomedical image segmentation,'' in \emph{Proc. of International Conference
  on Medical image computing and computer-assisted intervention (MICCAI)},
  2015, pp. 234--241.

\bibitem{abadi2016tensorflow}
M.~Abadi \emph{et~al.}, ``Tensorflow: A system for large-scale machine
  learning,'' in \emph{Proc. of USENIX Symposium on Operating Systems Design
  and Implementation (OSDI)}, 2016, pp. 265--283.

\bibitem{dozat2016incorporating}
T.~Dozat, ``Incorporating nesterov momentum into adam,'' in \emph{Proc. of
  International Conference on Learning Representations}, 2016.

\end{thebibliography}

\ifCLASSOPTIONcaptionsoff
  \newpage
\fi
\vfill
\end{document}